\documentclass[iop,apj,appendixfloats]{emulateapj}
\usepackage{apjfonts}

\gdef\cat{Ca\,II\,$\lambda$8498,8542,8662}
\gdef\nai{Na\,I\,$\lambda$8183,8195}
\gdef\wf{FeH\,$\lambda$9916}
\gdef\kms{km\,s$^{-1}$}
\gdef\msun{$M_{\odot}$}

\lefthead{van Dokkum \& Conroy}
\righthead{The Stellar IMF in Early-Type Galaxies. I. Data}
\slugcomment{Submitted to the Astrophysical Journal}
\begin{document}

\title{The Stellar Initial Mass Function in Early-Type Galaxies From
Absorption Line Spectroscopy. I. Data and Empirical Trends}

\author{Pieter G.\ van Dokkum\altaffilmark{1} \&
Charlie Conroy\altaffilmark{2,3}
}

\altaffiltext{1}
{Department of Astronomy, Yale University, New Haven, CT, USA}
\altaffiltext{2}
{Department of Astronomy \& Astrophysics, University of California,
Santa Cruz, CA, USA}
\altaffiltext{3}
{Harvard-Smithsonian Center for Astrophysics,
Cambridge, MA, USA}

\begin{abstract}

The strength of gravity-sensitive absorption lines
in the integrated light of old stellar populations
is one of the few direct
probes of the 
stellar initial mass function (IMF) outside of the Milky Way.
Owing to the advent of fully depleted
CCDs with little or no fringing
it has recently become possible to obtain
accurate measurements of these features.
Here we present 
spectra covering the wavelength ranges
$0.35\,\mu{\rm m} - 0.55\,\mu{\rm m}$ and
$0.72\,\mu{\rm m} - 1.03\,\mu{\rm m}$
for the bulge of M31 and
34 early-type galaxies from the SAURON sample,
obtained with the Low Resolution Imaging Spectrometer on
Keck. The signal-to-noise ratio is $\gtrsim 200$\,\AA$^{-1}$
out to $1\,\mu$m, which is sufficient to measure gravity-sensitive
features for individual galaxies and to determine how they depend
on other properties of the galaxies.
Combining the new data with previously obtained
spectra for globular clusters in M31 and the most massive elliptical
galaxies in the Virgo cluster
we find that the dwarf-sensitive \nai\ doublet
and the \wf\ Wing-Ford band increase systematically with velocity
dispersion, while the giant-sensitive
\cat\ triplet decreases with dispersion. These
trends are consistent with a varying IMF, such that galaxies with deeper
potential wells have more dwarf-enriched mass functions. In
a companion paper (Conroy \& van Dokkum 2012) we use a comprehensive
stellar population synthesis model to
demonstrate that IMF effects can be separated from age and abundance
variations and quantify the IMF variation among early-type galaxies.

\end{abstract}


\section{Introduction}

The form of the stellar initial mass function (IMF) is of fundamental
importance for many areas of astrophysics and one of the largest
uncertainties in the interpretation of the integrated light of
stellar populations.
The IMF is reasonably well constrained in the disk of the Milky Way
as stars can be counted more or less directly.
For the past decade the consensus
has been that the Milky Way IMF is a powerlaw with a logarithmic slope
of $\sim 2.3$ at
$M \gtrsim 1$\,\msun, with a gradual turnover at lower masses
(see, e.g., {Kroupa} 2001; {Chabrier} 2003). This turnover can
be interpreted as a characteristic mass: in the Milky Way disk, the
formation of stars with masses of a few tenths of the mass
of the Sun is apparently favored over the formation of lower and
higher mass stars.
This departure from a powerlaw
is important, as most of the stellar mass- and
number density is in the form of low mass stars. As a result,
apparently subtle changes in the form of the low mass IMF significantly
alter the mass-to-light ($M/L$) ratios of galaxies.
As an example, for the same
total luminosity, a {Salpeter} (1955) IMF
with a constant slope of $2.3$ down to $M=0.1$\,\msun\ implies a $1.6\times$
higher stellar mass than a {Chabrier} (2003) IMF.

It seems likely that the IMF in other present-day
spiral galaxies is similar to that in the Milky Way disk, but
that does not mean that the IMF has the same form in all galaxies and
at all epochs. In particular, the most
massive elliptical galaxies have had
very different star formation histories than spiral galaxies. Their
central
regions  are thought to have formed in short-lived,
highly dissipative events at high redshift
(e.g., {Naab} {et~al.} 2007; {Hopkins} {et~al.} 2009; {Kormendy} {et~al.} 2009; {van Dokkum} {et~al.} 2010; {Oser} {et~al.} 2010).
Densities,
temperatures, turbulent velocities, and the dust and metal content
were almost certainly different from conditions in the present-day
Milky Way, which may well have led to a different characteristic
stellar mass
(e.g., {Padoan} \& {Nordlund} 2002; {Bate}, {Bonnell}, \& {Bromm} 2003; {Larson} 2005; {Krumholz} 2011; {Myers} {et~al.} 2011).

Motivated by these and other
arguments, several recent studies have attempted to
observationally
measure or constrain the form of the IMF in elliptical galaxies or
their progenitors. In 2008, several papers argued that the IMF may have
been ``bottom-light'' (dwarf-deficient)
at early times, with a higher characteristic mass than
the Milky Way IMF. {van Dokkum} (2008) [vD08] used the ratio of luminosity
evolution to color evolution of massive galaxies in clusters to
constrain the IMF, a test first proposed by {Tinsley} (1980).
{Dav{\'e}} (2008) found that the specific star formation rates of
galaxies at $z\sim 2$ are difficult to explain in the context
of galaxy formation models unless the characteristic mass was
higher than today. {Wilkins}, {Trentham}, \&  {Hopkins} (2008), following {Fardal} {et~al.} (2007),
argued that the $z=0$ stellar mass density is lower than the integral
of the cosmic star formation history, unless star formation estimates
at high redshift overestimate the formation rate of low mass stars.

Short of counting individual stars,
the most direct way to constrain the low mass IMF is to detect
and quantify the light emitted by dwarf stars. As has been known
for a long time, this is possible
thanks to gravity-sensitive absorption features whose strengths
are different in dwarfs and giants
(see, e.g., {Spinrad} 1962; {Cohen} 1978; {Carter}, {Visvanathan}, \&  {Pickles} 1986; {Couture} \& {Hardy} 1993; {Conroy} \& {van Dokkum} 2012).
The strongest dwarf-sensitive features  are
the \nai\ doublet (e.g., {Faber} \& {French} 1980; {Schiavon} {et~al.} 1997a)
and the \wf\ Wing-Ford band (e.g., {Wing} \& {Ford} 1969; {Schiavon}, {Barbuy}, \&  {Singh} 1997b);
the strongest
giant-sensitive feature (in the optical) is the \cat\ triplet
(e.g., {Cenarro} {et~al.} 2003). This work is technically challenging
as dwarfs contribute only 5\,\% -- 10\,\% of
the integrated light of stellar populations. Therefore, a 30\,\%
absorption feature in the spectra of dwarf stars has a depth of only
a few percent in integrated light. Detecting IMF variations therefore
requires line measurements with exquisite accuracy ($\lesssim 0.3$\,\%)
in spectral regions that are plagued by strong sky emission  and (typically)
poor detector performance.

Owing to the advent of fully depleted, high resistivity CCDs it has
recently become
possible to measure absorption lines in the far red with the required
accuracy to detect variations in the dwarf-to-giant ratio.
Using the upgraded red arm of the Low Resolution Imaging Spectrograph
on Keck  (LRIS; {Oke} {et~al.} 1995; {Rockosi} {et~al.} 2010)
we found that massive elliptical galaxies in
the Virgo and Coma clusters have enhanced Na\,I and Wing-Ford
band absorption
compared to metal-rich globular clusters and
to expectations from stellar population synthesis models,
indicating that
the IMF in these galaxies is ``bottom-heavy'' with respect to that
of the Milky Way
({van Dokkum} \& {Conroy} 2010, 2011; {Conroy} \& {van Dokkum} 2012).
This result is consistent with constraints
on the masses of early-type galaxies as derived
from stellar dynamics and lensing
(e.g., {Treu} {et~al.} 2010; {Auger} {et~al.} 2010; {Spiniello} {et~al.} 2011; {Thomas} {et~al.} 2011; {Dutton}, {Mendel}, \& {Simard} 2012; {Spiniello} {et~al.} 2012; {Cappellari} {et~al.} 2012). However, it 
is opposite to the conclusions from the 2008 studies.

The main uncertainties in our initial study ({van Dokkum} \& {Conroy} 2010) are the
small sample size (four galaxies in Virgo with Na\,I and Wing-Ford measurements
and four galaxies in Coma with Na\,I measurements), the relatively low
signal-to-noise ratio (S/N) of the spectra, and the fact that
our modeling did not allow for abundance
variations of individual elements.
As discussed in detail in {Conroy} \& {van Dokkum} (2012) abundance variations can
be separated from IMF variations by analyzing different absorption
lines of the same element. As an example, the Wing-Ford band depends on the
IMF but also on [Fe/H], and by comparing the strength of
the Wing-Ford band to other iron lines the two variables can be separated.
This approach requires high quality data and a stellar
population synthesis model that allows simultaneous fitting of individual
elemental abundances, the IMF, and other stellar population parameters.

In the present paper
we describe newly obtained
high S/N Keck spectroscopy of 34 early-type galaxies
spanning a large range in velocity dispersion and abundance patterns.
We show that dwarf- and giant-sensitive absorption lines can be
measured accurately for individual galaxies, and compare the measurements
to the most massive Virgo ellipticals and to
M31 globular clusters.
In an Appendix we re-assess the 
results of vD08 in the context of recent results by
{van der Wel} {et~al.} (2008) and {Holden} {et~al.} (2010), as well as the {Conroy} \& {van Dokkum} (2012)
stellar population synthesis models.
In a companion paper (Conroy \& van Dokkum 2012b [paper II])
we fit the new Keck spectra with the stellar population
synthesis model of {Conroy} \& {van Dokkum} (2012) to quantify the
IMF variation.

\section{Sample and Observations}

\subsection{Sample Selection}

The primary sample comprises a subset of the early-type
galaxies observed in the SAURON
survey ({Bacon} {et~al.} 2001; {de Zeeuw} {et~al.} 2002).
Specifically, we observed 34 of the 48 E/S0 galaxies
listed in Table 1 of {Kuntschner} {et~al.} (2010). 
As discussed in {de Zeeuw} {et~al.} (2002) the SAURON parent
sample is not complete but
was selected to span a large range in ellipticity
and absolute magnitude. We could not observe the entire
{Kuntschner} {et~al.} (2010) sample because of visibility
constraints
and the fact that we only had a single night
of LRIS time. The 14 galaxies that were excluded are
NGC\,3032, NGC\,3156, NGC\,3489, NGC\,4150,
NGC\,4526, NGC\,4550, and NGC\,5831, because
we gave preference to galaxies with ages $\geq 9$\,Gyr;
NGC\,4374, NGC\,4387, NGC\,4477, and NGC\,5198, as 
we gave preference to
galaxies with metallicity $0.05<Z<0.20$ and $\alpha$-enhancement
$0.15<[\alpha/{\rm Fe}]<0.25$ in that same part of the sky;
and NGC\,5982, NGC\,7332, and NGC\,7457, as they were not
observable in January.

\begin{figure}[htbp]
\epsfxsize=8.5cm
\epsffile{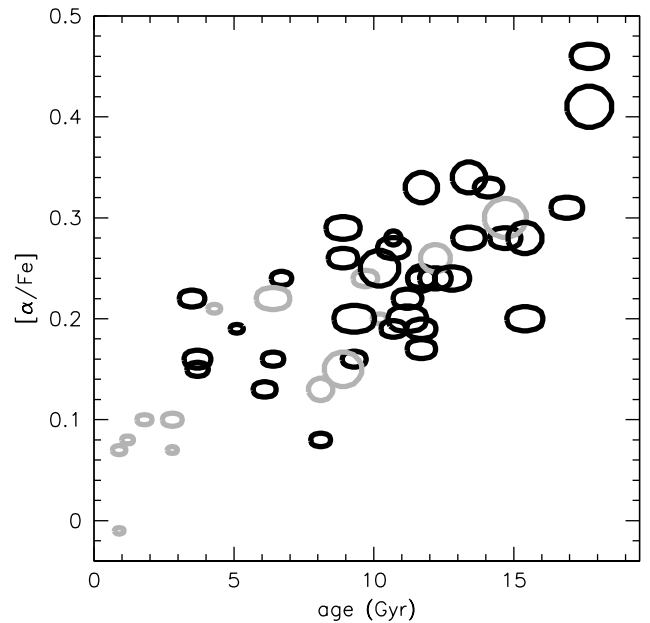}
\caption{\small
Sample selection. The points show the full SAURON sample of
Kuntschner et al.\ (2010) in the plane of
$\alpha$-enhancement
versus age within $r_e/8$. The symbol size scales with the
velocity dispersion. Circles denote slow rotators and ellipses
denote fast rotators. Black points were observed with LRIS;
grey points (mostly young galaxies with
low velocity dispersions) were not observed.
\label{sel.fig}}
\end{figure}

The sample is compared to the SAURON parent sample in Fig.\
\ref{sel.fig}. The LRIS sample discussed in this paper comprises
most of the SAURON sample, with an intentional bias
against the youngest
galaxies (which tend to be fast rotators
with low dispersions and 
low $\alpha$-enhancements). Using the derived quantities of
{Kuntschner} {et~al.} (2010) for $r<r_e/8$, the galaxies in
the LRIS sample have a median age of
11.2 Gyr, a median metallicity [Z/H] of 0.08, and a median
$\alpha$-enhancement of 0.24. It contains 7 slow rotators and 27
fast rotators, as defined by {Emsellem} {et~al.} (2007).

In addition to the SAURON galaxies we observed
the central regions of the bulge of M31. M31
has played an important if somewhat confusing role in the decades-long
quest to constrain the low mass end of the IMF through absorption line
spectroscopy. {Spinrad} \& {Taylor} (1971) and {Faber} \& {French} (1980) suggested that the
nuclear regions have a very large population of low mass stars, largely
based on anomalously strong \nai\ absorption. However, 
other studies have shown that the effects of
metallicity complicate the interpretation
(e.g., {Cohen} 1978; {Carter} {et~al.} 1986).

\subsection{Observing Strategy}
\label{strategy.sec}

The galaxies were observed on the night of
January 21, 2012 with LRIS
on the Keck I telescope.
The 680\,nm dichroic was used to split the light into the blue and red
arms. In the blue arm the 600\,l\,mm$^{-1}$ grism, blazed at
4000\,\AA, gave a spectral coverage of 3000\,\AA\ -- 5600\,\AA.
In the red arm the 600\,l\,mm$^{-1}$ grating blazed at 10,000\,\AA\
was set to cover the wavelength range 7100\,\AA\ -- 10,400\,\AA.
We used a relatively narrow ($0\farcs 7$)
slit to maximize the spectral resolution.
This
is not important for resolving lines in the galaxy spectra as they
all have $\sigma \gtrsim 100$\,\kms, but very helpful when correcting
for sky emission and absorption. The spectral resolution
$\sigma_{\rm instr}$, as measured
from sky emission lines, is $\approx 60$\,\kms\ in the blue arm
(at $5500$\,\AA) and ranges
from $\approx 65$\,\kms\ at 7200\,\AA\ to
$\approx 45$\,\kms\ at $9500$\,\AA\ in the red arm.
The red detector was binned by a factor two in the spatial direction
to reduce read-out time. After applying the same
binning (in software) to the data from the blue detector the
pixel scales are identical at $0\farcs 27$.

Each galaxy was observed for 540\,s, split into three 
180\,s exposures. The telescope was moved along
the slit between exposures,
such that each galaxy was observed in two  positions on
one of the two detectors and in one position on the other
detector. The slit was always positioned along the minor axis
of the galaxy to
minimize galaxy light near the edges of the slit and facilitate sky
subtraction. The white dwarf GD\,153
(see {Bohlin} 1996) 
was observed to correct for
the wavelength variation in the detector and instrument response.
At the beginning and end of the night arc lamp exposures
were obtained for wavelength calibration. 
Conditions were clear; we note that our initial data for
this project ({van Dokkum} \& {Conroy} 2010) were taken through clouds and therefore
had significantly lower S/N ratio than the data described here.

\section{Data Reduction}

\subsection{Overview}

Although the LRIS long slit has a length of $168\arcsec$ the effective
slit length is much shorter. The reason is that the LRIS blue and red
CCDs are both mosaics, and the slit is projected onto two independent
detectors. On the blue side, $75\arcsec$ is imaged on
one detector, $14\arcsec$ falls in a gap between two detectors, and
$79\arcsec$ falls on another detector. The detectors have slightly different
characteristics, and due to optical distortions and flexure the relation
between pixel location and wavelength needs to be determined independently
for each $\sim$\,half of the slit. As the flexure varies with the
position of the telescope,
each 540\,s sequence effectively comprises 12 independent exposures:
2 detectors $\times$ 2 arms $\times$ 3 dither positions.
Another
consequence of the somewhat peculiar slit geometry
is that the galaxies often
extend to the edges of the slit halves, complicating the sky subtraction.

Within these constraints the data reduction followed fairly
standard procedures: bias subtraction, using the overscan regions;
correction for s-distortion; wavelength calibration, using a combination
of arc lamps and the location of sky emission lines; subtraction of
a 2D model of the sky lines; cosmic ray identification
and combination of the individual science exposures in a sequence;
extraction of one-dimensional spectra,
mimicking a circular aperture of $r = r_e/8$; correction for
detector and instrument response; and correction for atmospheric
absorption. This last step is one of the most critical, as 
a near-perfect correction is required for our purposes. The steps
are detailed below.
The red and blue
spectra were treated in the same way, unless noted otherwise.

\subsection{Distortion Correction and Wavelength Calibration}
\label{lamcor.sec}

After bias subtraction the spectra were placed on an undistorted output
grid that is linear in the wavelength and spatial axes. 
The s-distortion was mapped by fitting the position of the galaxy
in the spatial direction with a Gaussian at 50-pixel intervals, and then 
fitting a 3$^{\rm d}$-order polynomial to the measured positions. 
A two-dimensional wavelength solution was obtained from arc lamp exposures,
by fitting 3$^{\rm d}$-order polynomials in the wavelength direction and the
spatial direction. The median residual
is typically $\approx 0.2$\,\AA; higher order polynomials did not improve
the fit.
These arc lamp solution capture the tilt of the sky lines and the distortion in
the wavelength direction, but flexure in the spectrograph
causes offsets of typically $\approx 2$\,\AA\ between
the science exposures and arc lamps. 
Bright sky emission lines were used to find the zero-order correction
for these differences between the arc lamps and the science exposures.
A small linear correction was applied after extraction of the spectra
(see \S\,\ref{skycor.sec}).

The spectra were mapped onto an output grid with 1\,\AA\ pixels in the
wavelength direction and $0\farcs 27$ pixels in the spatial direction,
using linear interpolation. This resampling method mostly conserves
the noise properties of the data and the sharp edges of cosmic rays, and
does not introduce significant aliasing in sky lines which would complicate
their subtraction.

\subsection{Sky Subtraction}

The subtraction of sky emission lines is complicated by the fact that
the galaxies cover a substantial fraction of the slit.
Sky lines were subtracted from the 2D spectra in several
steps, making use of the fact that
each of the two detectors has
at least one exposure in each sequence with negligible galaxy light
(see \S\,\ref{strategy.sec}). First, a 1D model of the sky was
created using these sky exposures, by median filtering in the
spatial direction. Next, the spatial variation in the sky
exposures was modeled using a low order polynomial. A 2D sky model
for each detector was generated by replicating the 1D model in the
spatial direction, weighted by the polynomial fit. This
(nearly) noise-free model of the sky emission
for each detector was then subtracted from the galaxy exposures.

This procedure is effective in removing sky emission without
affecting the galaxy light. However, residuals remain,
due to the variation in the intensity of sky lines on timescales
of a few minutes: the sky exposure is typically about
4 minutes removed in time from the galaxy exposures on the same
detector. These sky line residuals were removed by subtracting the
average of the two edges of the detector at each wavelength. This step
reduces the variation in the sky lines to the photon noise for most
galaxies, but it comes at a cost. For the largest galaxies there is
still detectable galaxy light at the edge of the slit, and by subtracting
this light we reduced the utility of the data for measuring gradients
in absorption features. Furthermore, if the gradients are strong the
subtraction alters the observed absorption line strengths. The maximum
effect on absorption lines occurs when the gradient is such that the
absorption feature
vanishes at the edge of the slit. In that case, the observed
absorption at radius $r$ will be increased by a factor
of $F_{\rm edge}/F_r$, with $F_{\rm edge}$ the
(subtracted) galaxy flux at the edge of the slit
and $F_r$ the galaxy flux at radius
$r$. The galaxy flux at the edge
is always $\ll 1$\,\% of the average flux in our extraction aperture;
this effect is therefore negligible in our analysis.

\subsection{Cosmic Ray Identification and Combination of Exposures}

Cosmic rays and other defects were identified in the following way.
First, a model for the galaxy light was created by taking the median
of the three individual exposures and then median filtering in the
spatial direction. After subtraction of this model, residual galaxy
light was removed by fitting and subtracting
a 7$^{\rm th}$-order polynomial in the spatial direction. The resulting residual
exposure contains noise and cosmic rays. Next, a 2D-model
of the total flux in each exposure was created from the 2D sky model
and the galaxy model. This total flux model $M$ was converted to a noise
model $N$ through $N = g^{-1}\sqrt{M \times g}$, with $g$ the gain.
A pixel was flagged as a cosmic ray when its flux in the residual
exposure was $7\times$ higher than the flux in the noise model. Due
to the linear resampling cosmic rays are slightly smoothed with respect
to the original exposures; to take this into account all pixels neighboring
cosmic ray pixels were also flagged.

The three exposures of each galaxy were summed to create a combined,
sky subtracted 2D output frame. Pixels affected by a cosmic ray in
one exposure were replaced by 1.5$\times$ the sum of the other
two exposures. In the rare cases where two exposures were affected by
a cosmic ray the pixel in the output frame is 3$\times$ the flux
in the unaffected exposure. The locations of pixels that were affected
by a cosmic ray in at least one of the three exposures are stored for
diagnostic purposes.

\subsection{Extraction of Spectra and Flux Calibration}

One-dimensional spectra were extracted from the 2D spectra by summing
the flux in the spatial direction. An extraction aperture of
$r_e/8$ was used; the effective radii were taken from
{Kuntschner} {et~al.} (2010) and corrected to the minor axis.
This aperture is also used by the
SAURON survey (along with larger apertures), allowing direct
comparisons to their results. A ``straight'' summation of the long slit
spectrum over the range $-r_e/8 < r < r_e/8$ would be weighted more
towards the center of the galaxy than the summation in a circular
aperture of SAURON. To mimic summation in a circular aperture with
radius $r=r_e/8$ we extracted the spectrum as follows:
\begin{equation}
F_{\lambda} = \displaystyle\sum\limits_{y=-1}^{1} F_{\lambda,y} +
\sum\limits_{y=-n}^{-2} -y F_{\lambda,y} + \sum\limits_{y=2}^n y F_{\lambda,y},
\label{extract.eq}
\end{equation}
with $y$ the pixel coordinate in the spatial direction (with the galaxy centered
in the middle of pixel zero) and $n$ the nearest integer
number of pixels corresponding to $r_e/8$. The first term in
Eq.\ \ref{extract.eq} is a straightforward
sum of the central three rows, corresponding to a
rectangular aperture of $0\farcs 81 \times 0\farcs 70$. The other
two terms extend the summation to $r_e/8$, weighting by the distance
from the central row. Weighted in this manner the spectra
can be
compared directly to the SAURON measurements, and represent
a larger fraction of the galaxy light. For an $r^{1/4}$ law,
a circular aperture of $r_e/8$ contains $\sim 10$\,\% of a galaxy's
total flux.

Next, the extracted spectra were calibrated using a response curve
that produces a flat spectrum for an object with constant $F_{\lambda}$.
No absolute calibration was attempted, but given
the rapid fall-off of the detector response at wavelengths
$\gtrsim 9600$\,\AA\ it is important to obtain a reasonably accurate
relative calibration as a function of wavelength. 
The response curve was created as follows.
First, a boxcar-smoothed
high S/N ratio halogen lamp spectrum was used to model the
small-scale variations in the response. This spectrum is shown by
the dashed line in Fig.\ \ref{resp.fig}. Next,
the extracted spectrum (corrected for atmospheric absorption;
see below) of the white dwarf GD\,153 was divided
by the halogen-derived response curve and then divided by the
Rayleigh-Jeans approximation of its spectrum
($F_{\lambda} \propto \lambda^{-4}$). The residual spectrum was
fitted by a low order polynomial. The final response
curve, indicated by the solid line in Fig.\ \ref{resp.fig},
was created by multiplying this polynomial
by the halogen-derived response curve. From a comparison of
the galaxy spectra to the stellar population synthesis models of
paper II we estimate that the relative
uncertainty in the calibration as a function of wavelength
is $\lesssim 5$\,\% on $\sim 1000$\,\AA\ scales.

\begin{figure}[htbp]
\epsfxsize=8.5cm
\epsffile{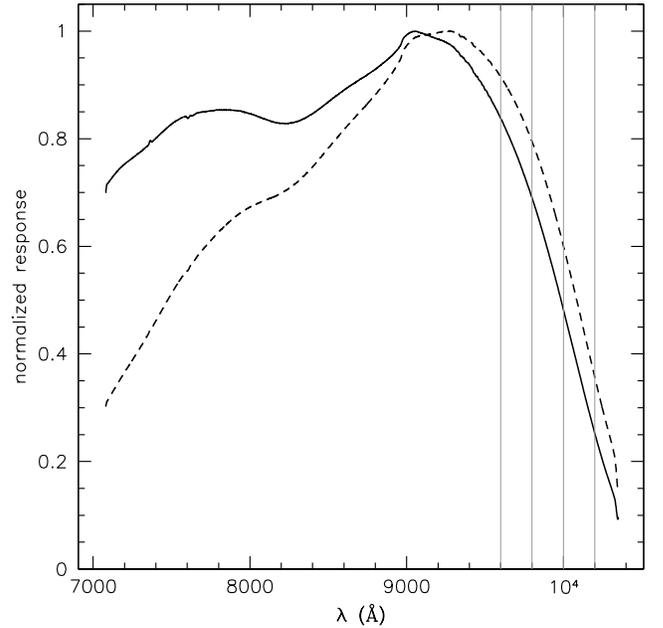}
\caption{\small
Response curve that was used to calibrate the red spectra (solid line),
as determined from the white dwarf GD\,153. Division by the response
curve produces a flat spectrum for an object with constant $F_{\lambda}$.
The dashed line shows a high S/N halogen lamp spectrum which
was used to capture the small-scale variation in the response.
\label{resp.fig}}
\end{figure}

\subsection{Correction for Atmospheric Absorption}

\begin{figure*}[htb]
\epsfxsize=15.5cm
\epsffile{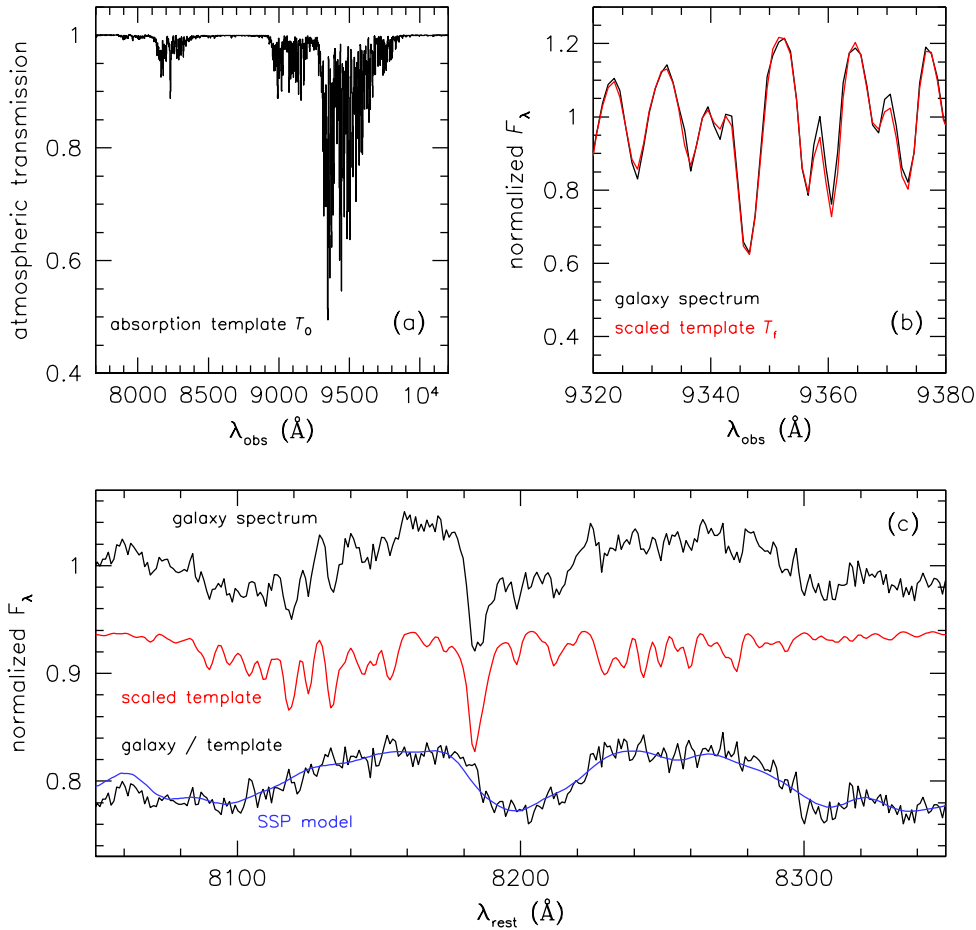}
\caption{\small
Illustration of our
correction for atmospheric absorption, for
NGC 5846. {\em (a)} Template atmospheric absorption spectrum.
{\em (b)} Galaxy spectrum (black) and template absorption spectrum
(red) in the $\sim 9350$\,\AA\ region, where the absorption is
strong. The template is scaled to match the observed galaxy
spectrum in this wavelength region.
{\em (c)} Galaxy spectrum before and after division by
the scaled absorption template, near the
Na\,{\sc i} doublet. The blue line shows the best-fitting
stellar population synthesis model from paper II.
\label{absorb.fig}}
\end{figure*}

As we aim to measure weak stellar
absorption lines with high accuracy it is
crucial to correct for absorption 
in our own atmosphere. The strongest atmospheric absorption features
in the optical are the ``A'' and ``B'' bands of O$_2$ at
$\sim 6870$\,\AA\ and $\sim 7600$\,\AA\ respectively. However, the
weaker but numerous H$_2$O bands in the
regions 8100\,\AA\ -- 8400\,\AA\
and 8900\,\AA\ -- 9800\,\AA\ are a more serious challenge
for our program,
as several key features (in particular the Na\,I doublet) fall in this
wavelength region.

The standard method to correct for atmospheric absorption is to observe
blue stars that are located near the science targets in the sky.
Typically a star is observed before and after the target, so that the varying
absorption can be interpolated to match the time and sky position
of the science observation (see, e.g., {Kriek} {et~al.} 2008). In practice
telluric standards are usually A stars, as O and B stars are rare and
typically not available in the general direction of the science target.
This procedure suffers from several drawbacks. A stars have strong Paschen 
lines at wavelengths $>8200$\,\AA, which need to be divided out before
the spectrum can be used to model the atmospheric absorption. They
also have weak metal lines which can introduce
systematic errors at the 0.5\,\% -- 1\,\% level. Finally, telluric
standards carry
significant overhead, particularly given the requirement that they
are observed before and after each science target.

Here we take a different approach, and correct for atmospheric absorption
by scaling a template spectrum to the observed absorption. The scaling
is parameterized by
\begin{equation}
T_f = 1 + f (T_0 - 1),
\end{equation}
with $f$ a scale factor and $T_0$ a template absorption spectrum. For
each galaxy the best-fitting value of $f$ is found by minimizing $|G - T_f|$,
with $G$ the galaxy spectrum. The fit is done over the wavelength range
9320\,\AA\ -- 9380\,\AA, as this region is dominated by strong
atmospheric H$_2$O lines and does not contain strong (redshifted)
galaxy absorption features. Prior to the minimization $G$ and $T_f$ are
divided by a 4$^{\rm th}$-order polynomial fit in the wavelength range
9250\,\AA\ -- 9650\,\AA. The
template $T_0$ is appropriate for Mauna Kea and smoothed
to the instrumental resolution.

The procedure (also explored in {Schiavon} 1998)
is illustrated in Fig.\ \ref{absorb.fig} for one of the
sample galaxies (NGC\,5846). The theoretical absorption template
is shown in panel (a). Panel (b) shows the detailed region of the
template around the strongest absorption lines, where the fit
is done to determine the best match to the galaxy spectrum.
The top spectrum in
panel (c) shows the significant effect of
atmospheric absorption lines on the region
around the \nai\ feature. After division by the scaled template
(red) the narrow sky lines are nearly perfectly removed. The
blue line is the best-fitting template from paper II. An
analysis of the residuals of these fits in regions affected by
sky absorption shows that the absorption correction is good to about
$5-10$\,\% {\em of the feature strength}. At our
spectral resolution the
strongest absorption feature in the 8000\,\AA\ -- 8500\,\AA\
range is $\sim 10$\,\%, which implies
that the largest residuals are $0.5-1$\,\%.
As the width of sky absorption lines
is typically $\sim 20$\,\% of the width
of galaxy absorption features, this results in a $0.1- 0.2$\,\%
uncertainty in the strength of a galaxy absorption line
that coincides with a
relatively strong sky absorption line.

\subsection{Optimizing the Wavelength Calibration}
\label{skycor.sec}

The standard wavelength calibration as described in \S\,\ref{lamcor.sec}
is correct to approximately $\pm 1$ pixel
($\pm 30$\,\kms) over the entire wavelength range. Although this
level of accuracy is sufficient for most purposes, it is
a source of uncertainty in the analysis in paper II.
The reason for this sensitivity to the wavelength calibration
is that in our methodology a template spectrum is directly fit
to the observed spectrum. As a result, a small error in the
wavelength calibration increases the $\chi^2$ value of the fit.
In the fitting this increase can be
partially compensated by changing the line strength
in the model, thus potentially
leading to erroneous abundances and other
fit parameters.

\begin{figure}[htbp]
\epsfxsize=8.5cm
\epsffile{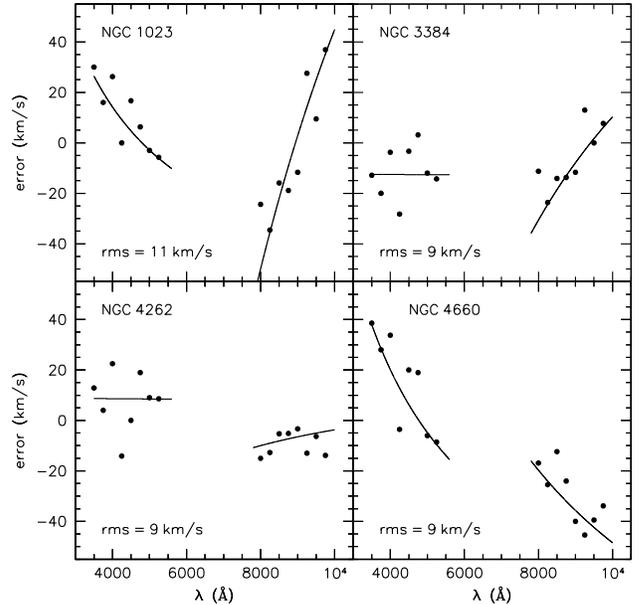}
\caption{\small
Errors in the wavelength calibration, resulting from wavelength-
and time-dependent
differences between the arc lamp solution and the galaxy spectra.
The points were determined from model fits in 250\,\AA\ wide
spectral regions;
the four  galaxies that are shown span the full range of variation in
the errors. Lines indicate linear fits (in wavelength) to the errors;
the residuals after this correction are $\sim 10$\,\kms.
\label{skycor.fig}}
\end{figure}

We optimized the wavelength calibration by fitting an
$\alpha$-enhanced, 13.5 Gyr old stellar population
synthesis model to the blue
and red spectra in narrow wavelength regions. The spectra were divided
in $\sim 20$ regions, each with a width
of 250\,\AA. The model was smoothed to the velocity dispersion of
the galaxies and fit to each of the regions, with velocity as the only
free parameter. The deviations from the average velocity, expressed
in \AA,
were fit with a linear function in wavelength. Each galaxy was fit
separately, and the red and blue spectra were treated independently.
The linear fits to the corrections were then
applied to the wavelengths of the extracted spectra.
The fit procedure is illustrated in Fig.\ \ref{skycor.fig}, for four
galaxies that exhibit the full range of corrections. 
The slope of the required corrections
broadly correlates with the time of night (and hence with the NGC number
of the galaxies), presumably because the true wavelength
calibration deviates more and more from the arc lamp solution.
After this correction the systematic errors are less than $10$\,\kms\
over the full wavelength range from
3500\,\AA\ to 10,000\,\AA\ for all galaxies.

\section{Extracted Spectra}

\begin{figure*}[p]
\epsfxsize=17cm
\epsffile{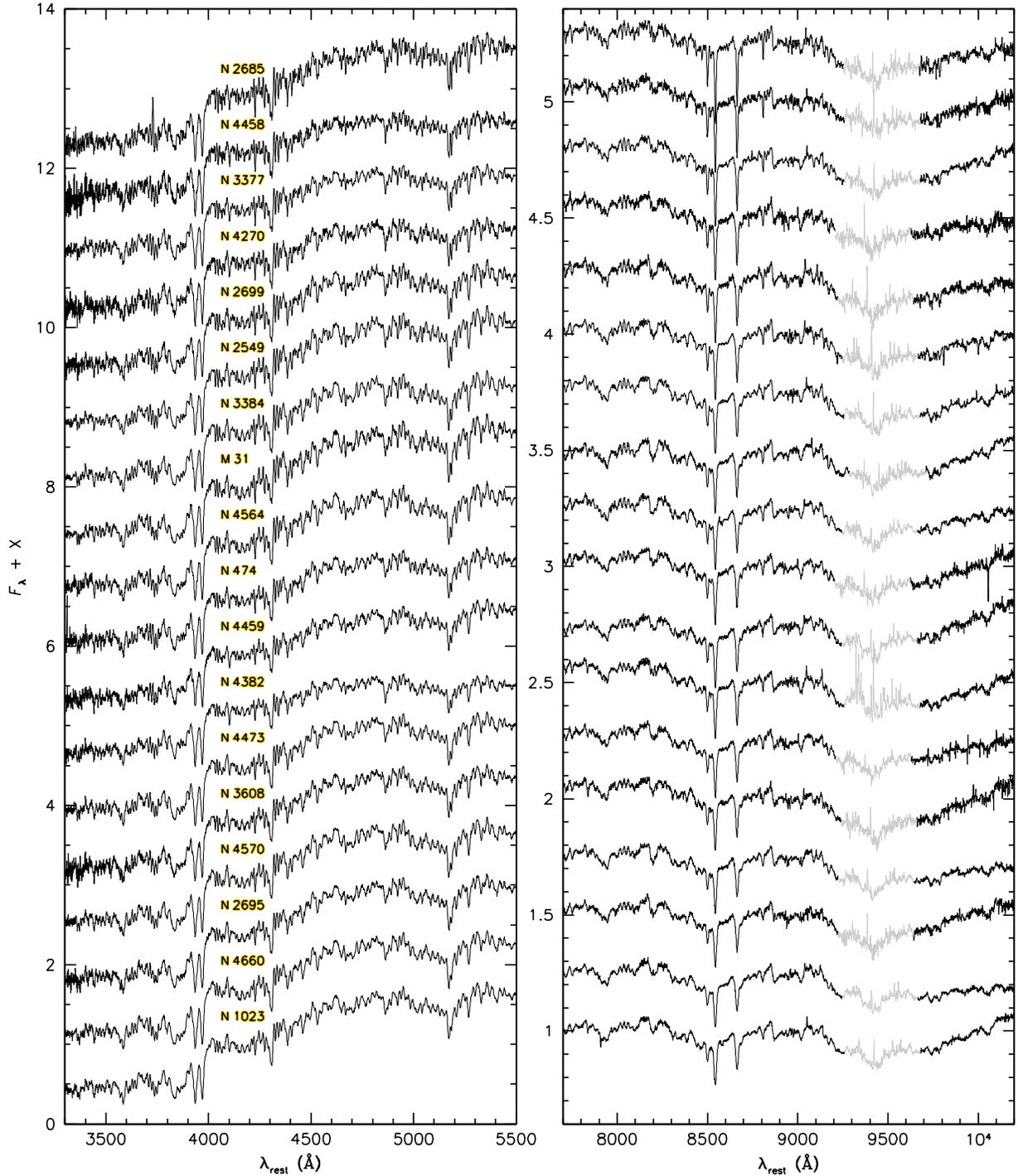}
\caption{\small
Extracted LRIS spectra in the rest-frame,
using a (circularized) $r<r_e/8$ aperture. Starting
at the  top galaxies are ordered by increasing velocity dispersion. The
spectra are normalized at 4050\,\AA\ (blue side) and 9050\,\AA\
(red side). The region of strong sky absorption around 9500\,\AA\ is shown
in light grey. The blue cutoff in the displayed red spectra is dictated by
the onset of the atmospheric B band at $\sim 7600$\,\AA. The
spectra are of high quality, and extend beyond $1\,\mu$m in the
rest-frame.
\label{spec.fig}}
\end{figure*}
\addtocounter{figure}{-1}

\begin{figure*}[p]
\epsfxsize=17cm
\epsffile{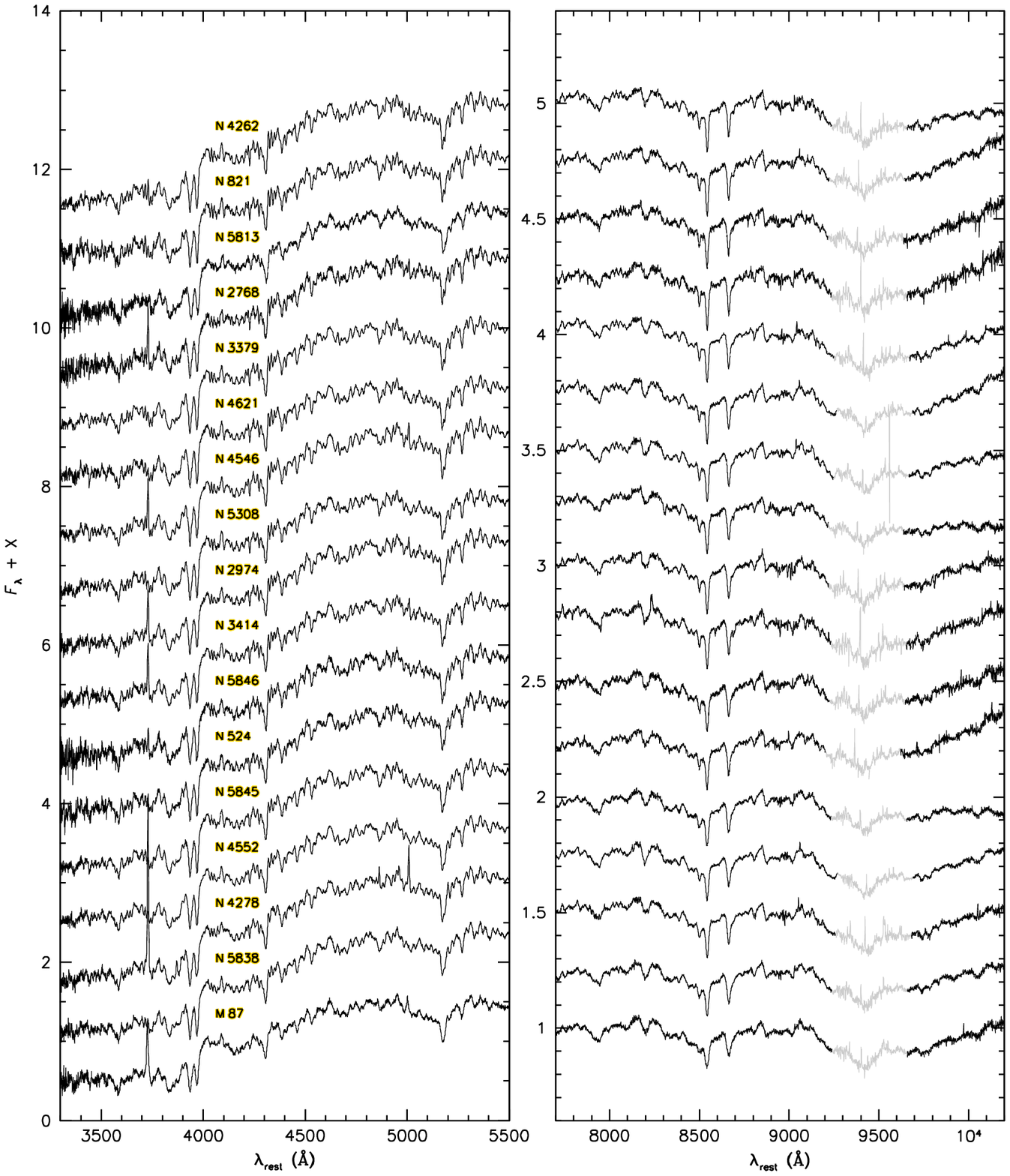}
\caption{\small
{\em (continued)}
}
\end{figure*}

The extracted spectra are shown in Fig.\ \ref{spec.fig}, ordered by increasing
velocity dispersion. The quality is generally high: essentially all the visible
features in the spectra (except in the far blue and far red) are spectral
lines, not noise. Nevertheless, it is clear that
some spectra have a higher S/N ratio than others. This variation largely stems
from the fact that each galaxy was observed for 540\,s irrespective
of its brightness. The region around 9500\,\AA\ suffers from strong sky
absorption (see Fig.\ \ref{absorb.fig}a), and is shown in light grey.

\begin{figure}[htbp]
\epsfxsize=8.5cm
\epsffile{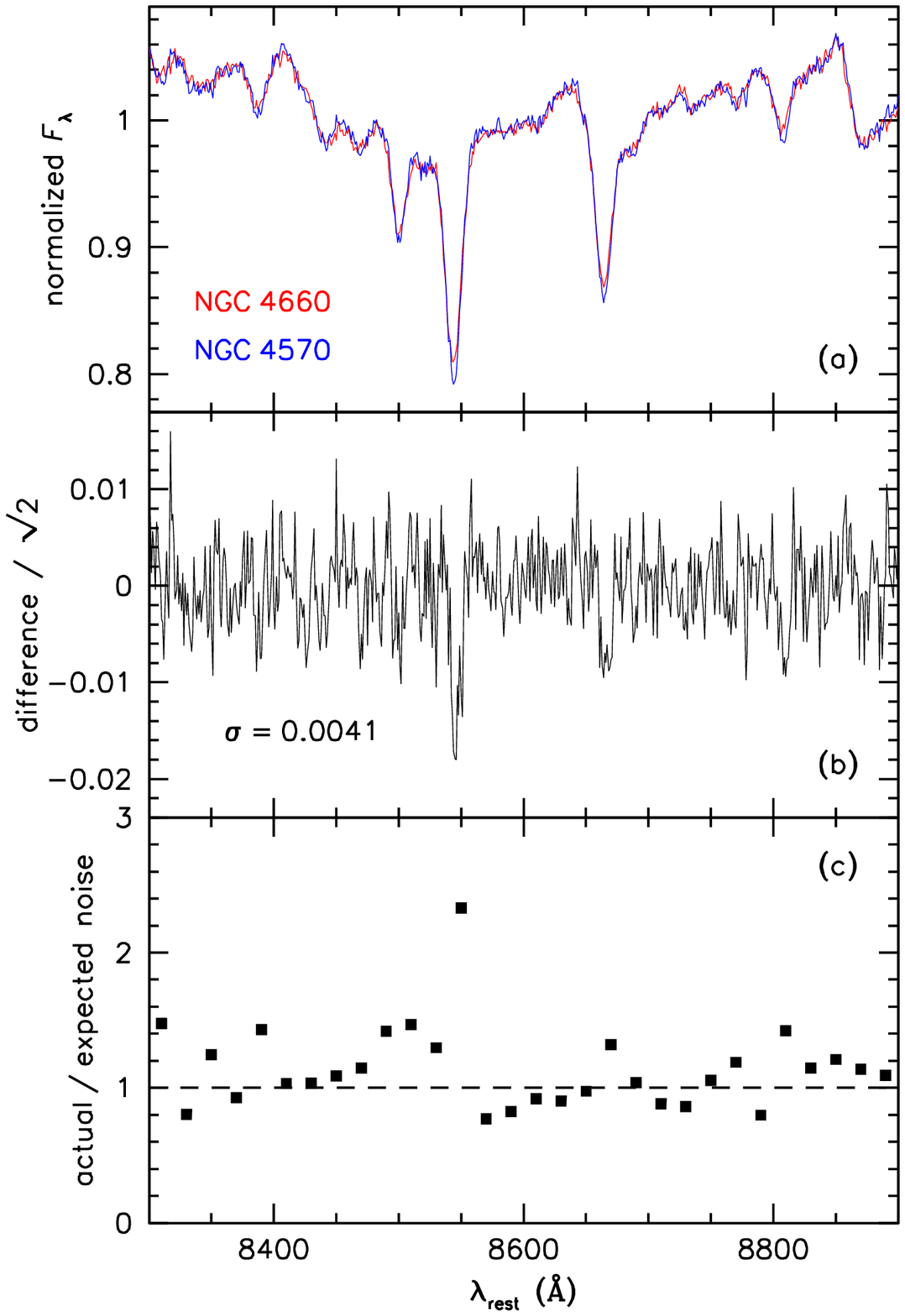}
\caption{\small
{\em (a)} Comparison of the spectra of
two galaxies with similar ages, abundances, and
velocity dispersions, in the
wavelength region near the
Ca\,II triplet. The spectra have a S/N ratio
of $\approx 250$\,\AA$^{-1}$, which is typical for our sample.
{\em (b)} Difference of the two spectra.
Apart from a slight difference in
Ca\,II absorption the variation between the spectra is very
small and appears mostly random.
The differences (divided
by $\sqrt{2}$) have a 68\,\% range of $\pm 0.4$\,\%. {\em (c)}
Observed scatter in {\em (b)} divided by the expected noise from
Poisson statistics. Away from strong absorption features the
observed differences between NGC\,4660 and NGC\,4570 are fully
consistent with the formal errors. This demonstrates that the
formal errors are a good approximation of the uncertainties
in the spectra.
\label{noise.fig}}
\end{figure}

For a correct interpretation of the model fits in paper II it is crucial
to have realistic estimates of the noise in the spectra. The
formal noise was determined
from Poisson statistics, taking the gain, the sky spectrum,
and the applied weighting (Eq.\ \ref{extract.eq}) into account.
It is difficult to test whether the actual noise corresponds to the
expected noise, as the observed variation in the spectra is almost
entirely due to the ``forest'' of weak absorption lines present in
the atmospheres of cool stars. We empirically determined the noise
properties of the spectra in the following way. We selected
two galaxies with very similar velocity dispersions, [Fe/H], and
[Mg/Fe] (as determined in paper II), and an average S/N ratio which is
typical for the full sample. The galaxies NGC\,4570 and NGC\,4660
satisfy all these critera. A small part of their spectra, centered
around the calcium triplet, is shown in
Fig.\ \ref{noise.fig}(a). As expected from the selection the
spectra are very similar, although NGC\,4570 has slightly stronger
absorption lines than NGC\,4660. 

The difference between the two normalized spectra is
shown in Fig.\ \ref{noise.fig}b.
The $1\sigma$ scatter in the difference spectrum is 0.0059\,\AA$^{-1}$,
as determined with the biweight estimator (e.g., {Beers}, {Flynn}, \& {Gebhardt} 1990).
Assuming each galaxy contributes equally to the scatter
this corresponds to $\sigma \approx 0.0041$\,\AA$^{-1}$ for each individual
galaxy. In Fig.\ \ref{noise.fig}c the scatter is calculated
in bins of 20\,\AA\ and divided by the average
formal error in the two spectra.
A value of 1 implies that the observed
differences between the spectra can be fully explained by Poisson
statistics. The median is 1.1 over the
displayed wavelength range and 1.2
over the entire 7800\,\AA\ --
10,200\,\AA\ range, which means that the formal errors are
a good approximation of the actual uncertainties in the
spectra.  Similar comparisons of other galaxy pairs
consistently show that the formal errors are reasonable, particularly
in regions away from strong sky emission or absorption lines.

\begin{figure}[htbp]
\epsfxsize=8.5cm
\epsffile{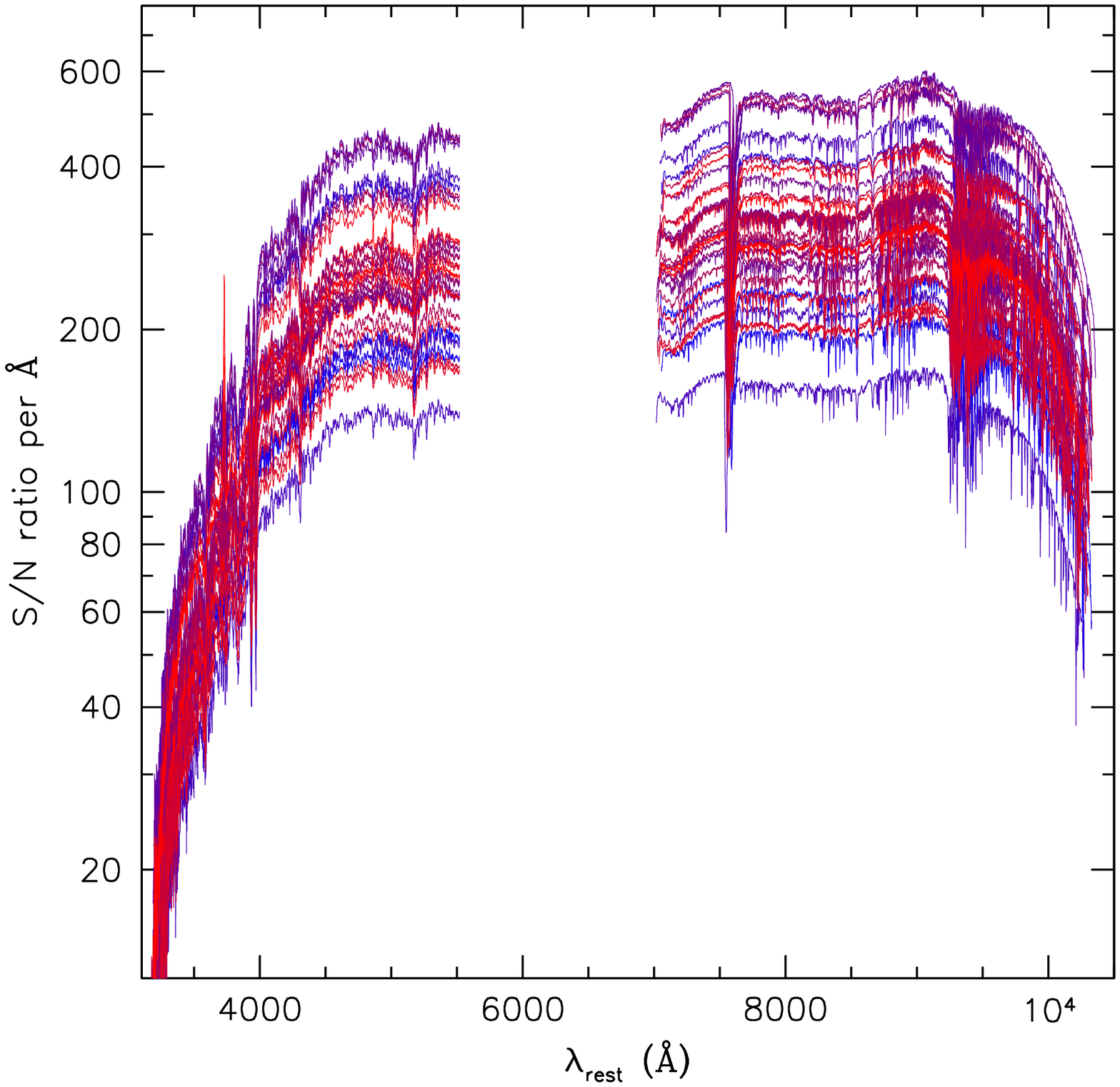}
\caption{\small
S/N ratio per rest-frame
\AA\ for all spectra. 
For individual galaxies the S/N is fairly uniform between $\sim 4500$\,\AA\
and $\sim 10,000$\,\AA, but there is
considerable variation between galaxies.
The spectra are color-coded according to their velocity dispersion,
from low = blue to high = red. There is no strong correlation between
S/N ratio and velocity dispersion.
\label{sn.fig}}
\end{figure}

Having verified that the formal S/N ratios are reasonable, we
show the S/N as a function of wavelength for all galaxies
in Fig.\ \ref{sn.fig}.
As expected
there is significant variation between galaxies and between
wavelengths. The S/N ratio ranges from
$\sim 60$ in the blue for the worst spectra to $\sim 500$ in the
red for the best spectra. The median S/N of all galaxies is 90\,\AA$^{-1}$
at 3800\,\AA, 256\,\AA$^{-1}$ at 5000\,\AA, 310\,\AA$^{-1}$ at
8500\,\AA, and 211\,\AA$^{-1}$ at 10,000\,\AA. There is no obvious
trend with velocity dispersion.

\begin{figure*}[htbp]
\epsfxsize=17cm
\epsffile{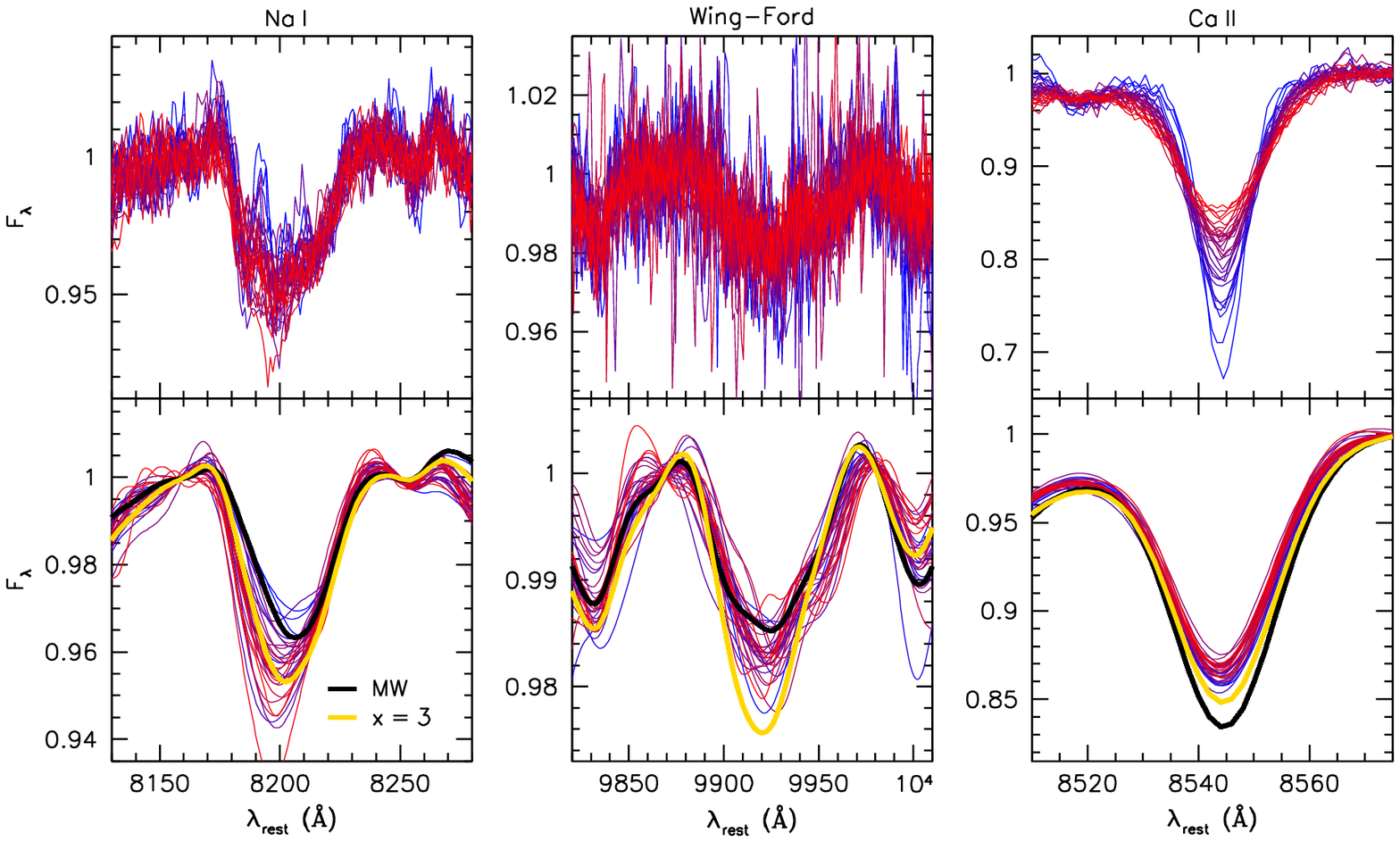}
\caption{\small
IMF-sensitive absorption features. Na\,I and FeH are strong in dwarf
stars and weak in giants; the calcium triplet is strong in giants and
weak in dwarfs. Top panels are at the original resolution; in the bottom
panels the spectra are smoothed to a common dispersion of 300\,\kms.
The spectra are color-coded by their velocity dispersion, going from
blue (low) to red (high). The black and yellow
lines show expectations for 13.5\,Gyr old,
$\alpha$-enhanced stellar populations with two different IMFs
(Milky Way and bottom-heavy). The dispersion-matched
spectra have sufficiently high S/N ratio to distinguish between
these predictions.
\label{feature.fig}}
\end{figure*}

\section{Analysis of IMF-Sensitive Features}

Here we investigate the observed
variation in the well-established IMF-sensitive
features \nai, the \cat\ triplet, and the \wf\ Wing-Ford band.
The Na\,I doublet and the Wing-Ford band are strong in dwarf stars and weak
in giants; conversely, the Ca\,II triplet is strong in giants and
weak in dwarfs. As a result, galaxies with
bottom-heavy IMFs are expected
to have stronger Na\,I, stronger Wing-Ford,
and weaker Ca\,II absorption than galaxies with bottom-light
IMFs (see {Couture} \& {Hardy} 1993; {Cenarro} {et~al.} 2003; {van Dokkum} \& {Conroy} 2010; {Conroy} \& {van Dokkum} 2012, and many other
studies).

\subsection{Dispersion Matching}

Prior to measuring the strength of absorption lines the galaxies have
to be smoothed to the same velocity dispersion. 
Velocity
dispersions of the individual galaxies were measured directly
from the extracted spectra, taking the instrumental resolution
and the resolution of the template into account (see paper II).
All galaxies except M87 were smoothed to a common resolution,
using a Gaussian of width 
\begin{equation}
\sigma_{s} = \sqrt{300^2 - (\sigma_*^2 + \sigma_{\rm instr}^2)},
\end{equation}
with $\sigma_*$ the stellar velocity dispersion and
$\sigma_{\rm instr}$ the instrumental resolution.
This smoothing has the potential to broaden localized sky line
residuals, thus ``contaminating'' the spectrum on 300\,km/s scales.
To prevent this, and to reduce the effect of sky line residuals on
measured absorption line indices,
pixels coinciding with strong sky lines were not taken into account in
the smoothing. This was done iteratively, in each iteration
replacing pixels in the unsmoothed spectrum that coincide
with sky lines by pixels from the smoothed spectrum
of the previous iteration.

\subsection{Na\,I, the Wing-Ford band, and Ca\,II}

The three IMF sensitive absorption lines are detected with
high significance in all galaxies, as
shown in Fig.\ \ref{feature.fig}.
The top panels show the spectra at their original spectral resolution;
in the bottom panels they are smoothed to a common velocity dispersion
of 300\,\kms. For clarity only one line of the calcium triplet is shown.
The
spectra were divided by a linear fit to the side-bands of the features
(see below).
The bulge of M31 and M87 are not shown; NGC\,3414 and NGC\,3608
were also excluded because they have unexplained noise peaks in their
Na\,I (NGC\,3414) and Wing-Ford (NGC\,3608) regions.

The black and yellow lines in Fig.\ \ref{feature.fig} illustrate the
IMF-sensitivity of these lines, using stellar population synthesis models
from {Conroy} \& {van Dokkum} (2012).
Both models have an age of 13.5\,Gyr,
a Solar iron abundance, and are $\alpha$-enhanced
with [$\alpha$/Fe]\,=\,0.2. The black model is for a {Chabrier} (2003)
IMF and the yellow model is for a bottom-heavy IMF with a logarithmic slope
$x=3$. The data span a similar range as these model predictions.
More to the point in the context of the present paper,
the data quality is sufficiently
high to measure the subtle differences in absorption line
strength expected from IMF variations.

\begin{figure}[htbp]
\epsfxsize=8.5cm
\epsffile{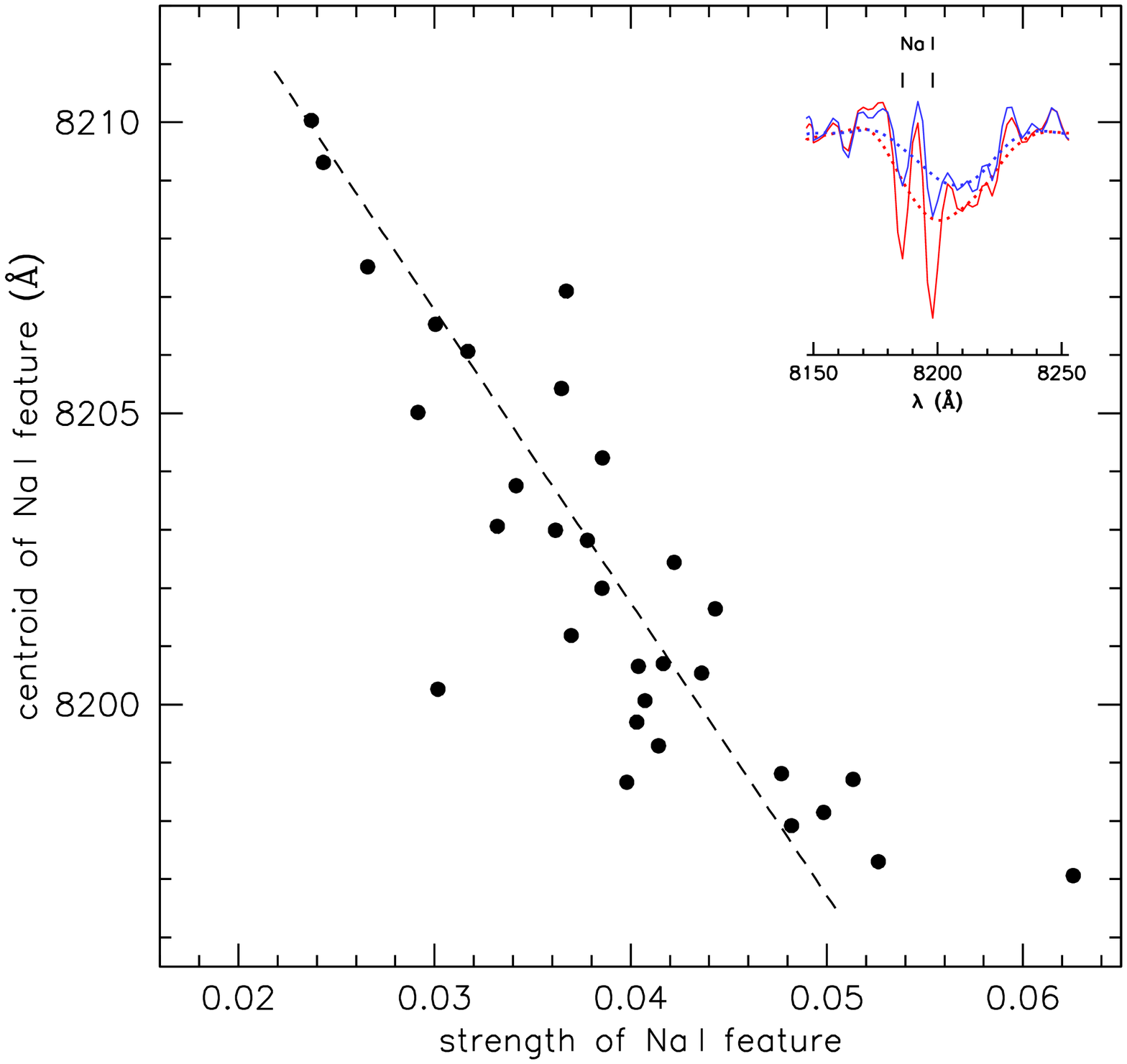}
\caption{\small
Correlation between the centroid of the Na\,I feature and its strength.
The inset shows the origin of the correlation. The blue line is for
a bottom-light IMF with weak Na\,I absorption and the red line is for
a bottom-heavy IMF with strong Na\,I absorption. The broken lines show
the same spectra at 300\,\kms\ resolution. At this resolution the
doublet is blended with a TiO bandhead at $\sim 8205$\,\AA, and
increased Na\,I absorption moves the centroid of the measured
absorption feature toward that of the bluer Na\,I doublet.
\label{centroid.fig}}
\end{figure}

As can be seen in the bottom left panel of Fig.\ \ref{feature.fig}
the line profile of the observed Na\,I feature changes with its depth:
its centroid is bluer when the absorption is stronger. We show this
relation between the strength of Na\,I and its centroid (here
simply defined as the wavelength of maximum absorption) explicitly
in Fig.\ \ref{centroid.fig}. The relation
arises because Na\,I
is a blend of
the \nai\ doublet and the TiO (0,2) bandhead
at a resolution of 300\,\kms\ (see {Schiavon} {et~al.} 1997a).
As the sodium absorption becomes stronger the centroid of the feature
shifts toward the Na\,I doublet and away from the TiO bandhead.
This effect is illustrated in the inset of Fig.\ \ref{centroid.fig},
which shows the effect of increasing the number of dwarf
stars (and hence the sodium doublet strength) on the measured
feature at 300\,\kms\ resolution. The existence of the tight relation
in Fig.\ \ref{centroid.fig} demonstrates that the observed
variation in the Na\,I feature strength is driven by variation
in Na\,I  absorption, not variation in the TiO band.

\begin{figure*}[htbp]
\epsfxsize=17cm
\epsffile{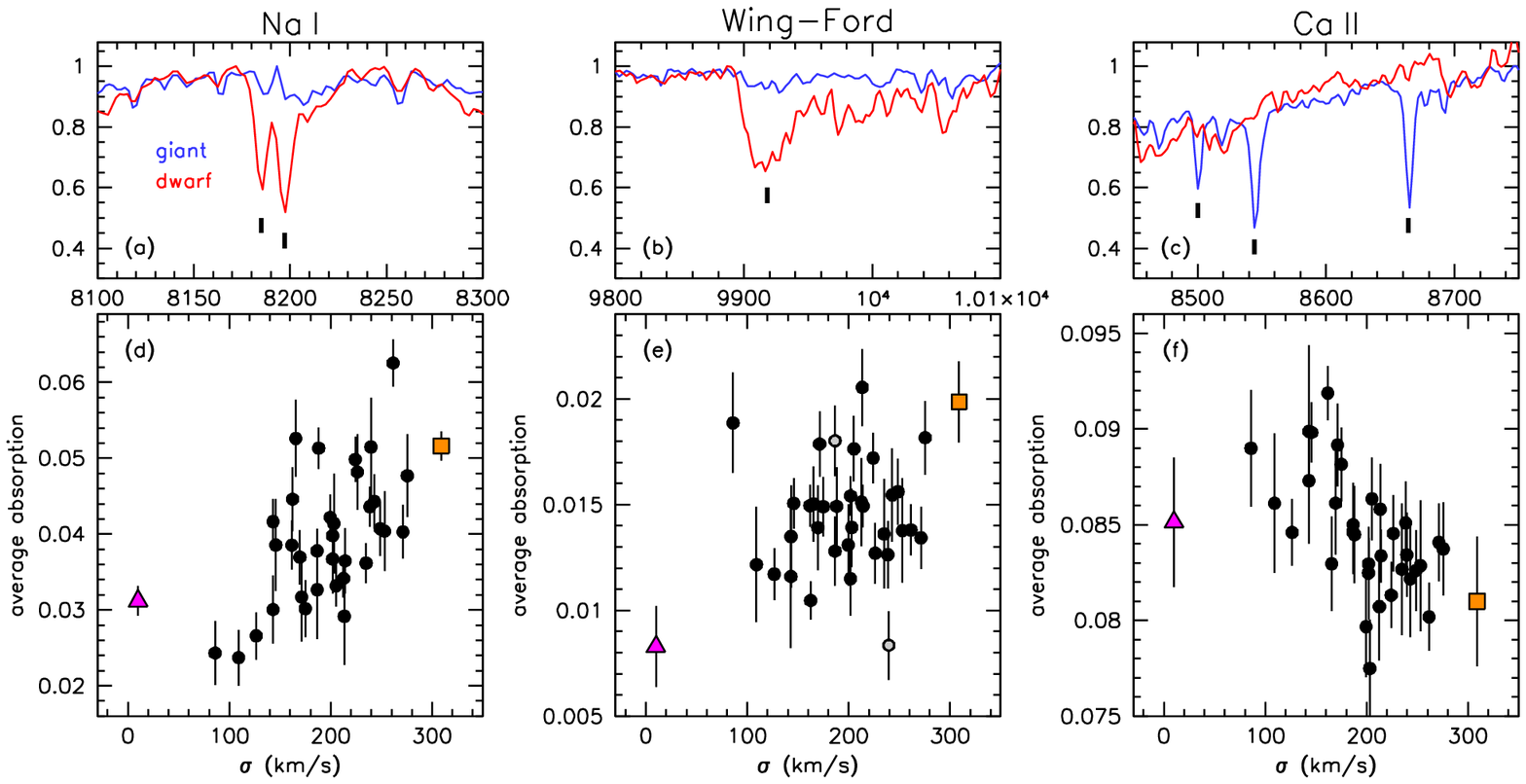}
\caption{\small
{\em (a-c):} The Na\,I doublet, the Wing-Ford band, and
the Ca\,II triplet in the M6 dwarf Gliese 406 (red) and the
M4 giant HD\,4408 (blue). For a larger dwarf-to-giant ratio
Na\,I and Wing-Ford are expected to be stronger and Ca\,II is
expected to be weaker. {\em (d-f):} The strength of these features
in integrated light, as a function of velocity dispersion. Black dots
are individual SAURON galaxies.
The purple triangle represents metal-rich
globular clusters in M31 (van Dokkum \& Conroy 2011), and 
the orange square is measured from the average spectrum
of high-dispersion
elliptical galaxies in the Virgo cluster (van Dokkum \& Conroy 2010).
Na\,I and Ca\,II show strong and opposing trends, consistent with
more bottom-heavy IMFs for galaxies with higher dispersions.
The relation between
the Wing-Ford band and $\sigma$ is only significant when the globular
clusters and the massive Virgo galaxies are included.
\label{index.fig}}
\end{figure*}

The relation between centroid and feature strength also provides
an empirical upper limit to the errors in
the line strength measurements.
Excluding the two most deviant points,
we find that the centroid of the feature predicts its
strength with an rms scatter of only 0.0033 (dashed line
in Fig.\ \ref{centroid.fig}). This is a strict upper
limit on the error
as errors in the centroid measurements,
variations in the TiO bandhead,  and variation
in other absorption features in
the central band or side bands all contribute to this scatter.
The tight relation has another interesting implication. As will be shown in
paper II the form of the IMF correlates
with the Na\,I strength. Therefore
one could, in principle, infer the IMF from a simple, model-independent
measurement: the centroid of the Na\,I feature at a
resolution of 300\,\kms.

\subsection{Correlations With Velocity Dispersion}

The spectra in Fig.\ \ref{feature.fig} are color-coded by their velocity
dispersion. In the top panels the dispersion is trivially related to the
width of the feature; this is particularly obvious for the calcium line
in the top right panel. In the bottom panels the spectra are smoothed
to the same dispersion, and yet trends with the galaxies' velocity
dispersion remain: galaxies with high velocity dispersions tend
to show strong Na\,I absorption and weak Ca triplet absorption.

We analyze these trends  by measuring the absorption line strengths
of the IMF-sensitive features. Such line strengths are useful
for highlighting trends of specific absorption features in the data.
However, we note here that line indices are notoriously difficult
to interpret quantitatively
as they rarely measure the abundance of a single element
in a straightforward way. Their central bands or side bands typically
contain faint lines of other elements (e.g., {Schiavon} {et~al.} 1997a; {Kelson} {et~al.} 2006);
they suffer from well-documented
degeneracies between age and abundance
(e.g., {Worthey} 1994); and multiple features of
the same element are typically
needed to differentiate abundance effects from IMF effects
({Conroy} \& {van Dokkum} 2012). For these reasons we do not use line indices
in paper II, where we quantify the IMF, but fit
the galaxy spectra directly with comprehensive
stellar population synthesis models.

The relation between the strength of IMF sensitive features and
velocity dispersion is shown in Fig.\ \ref{index.fig}.
The line strengths are defined as the average absorption over a
central band, with the continuum determined by interpolating between
two side bands. The errorbars are a combination of the formal
Poisson uncertainty (which dominates for the Wing-Ford band and Ca\,II)
and the uncertainty introduced by the atmospheric absorption correction
(which dominates for Na\,I).
For Na\,I and the Wing-Ford band
we use the same central and side band
definitions as in {van Dokkum} \& {Conroy} (2010).
For Ca\,II we use the definitions of {Conroy} \& {van Dokkum} (2012).
Also shown are measurements from stacked spectra of metal-rich
globular clusters in M31 from {van Dokkum} \& {Conroy} (2011) and of high velocity
dispersion elliptical galaxies in the Virgo cluster from {van Dokkum} \& {Conroy} (2010).

The strength of the two dwarf-sensitive features
(Na\,I and the Wing-Ford band)
systematically increases with velocity dispersion. M31 globular
clusters have the weakest absorption, the four Virgo ellipticals
from {van Dokkum} \& {Conroy} (2010) have the strongest absorption, and the SAURON
galaxies fall in between. By contrast, the strength of the
giant-sensitive Ca\,II triplet {\em decreases} with velocity dispersion,
as was found earlier by {Cenarro} {et~al.} (2003). These trends are
consistent with a systematically increasing dwarf contribution
with $\sigma$. We note, however, that the Wing-Ford band does not
show a significant correlation with $\sigma$ within the SAURON
sample (i.e., disregarding the globular clusters and the most massive
ellipticals). The correlation coefficient is positive (0.12) but not
significant. By contrast, the probability that the
(anti-)correlations of Na\,I and
Ca\,II with $\sigma$ are caused by chance are 0.3\,\% and 0.1\,\%
respectively.

\begin{figure}[htbp]
\epsfxsize=8.5cm
\epsffile{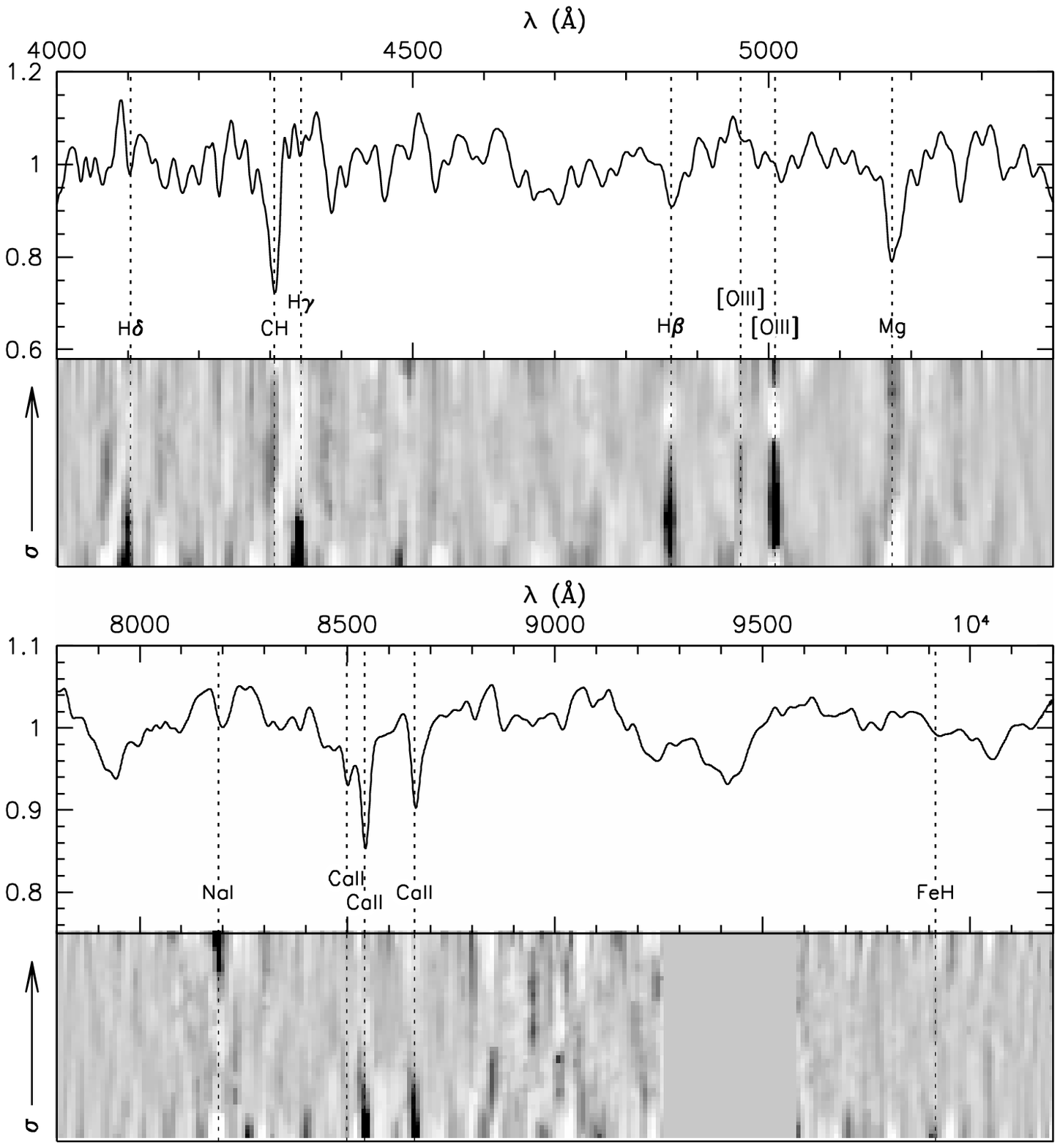}
\caption{\small
Variation in the spectra of the SAURON galaxies.
The top panels show
the average de-redshifted
spectrum of all
sample galaxies, smoothed to a common velocity dispersion of
300\,\kms\ and divided by a 5th order polynomial. The greyscale
shows the differences
between individual galaxy spectra and this averaged spectrum,
after smoothing and median filtering.
For clarity the
variation in the red is increased by
a factor of three compared to the variation in the blue.
Of all spectral lines at $\lambda>7800$\,\AA\
the IMF-sensitive
Na\,I and Ca\,II features show the strongest trends with
$\sigma$.
\label{var.fig}}
\end{figure}

The trends within the SAURON sample are graphically illustrated in
Fig.\ \ref{var.fig}, which shows the variation in the galaxy
spectra  ordered by velocity dispersion.
It is remarkable that
the IMF-sensitive Na\,I and Ca\,II features show the strongest
variation of any lines in the red. 
Figure \ref{var.fig} also highlights the well-known
fact that many other
spectral features show systematic trends with $\sigma$:
most notably the Balmer lines and the [OIII] emission lines, but
also Mg\,$\lambda 5270$ and a host of other metal lines
(see, e.g., {Trager} {et~al.} 2000b, and many other studies). 

\section{Summary and Conclusions}

In this paper we presented deep spectroscopy of a sample of early-type
galaxies in the nearby Universe, obtained with LRIS
on Keck I.
The spectra are weighted in such a way that they are
representative for a circular aperture of radius $r = r_e/8$. Owing to
the fully depleted LBNL detectors in the red arm of LRIS the S/N ratio
of the spectra is high all the way to $\sim 1\,\mu$m. The high S/N ratio
and the absense of fringing make it possible to measure absorption lines
with $<0.5$\,\% uncertainty in the far red. The reduced
spectra are 
available upon request.

The analysis in the present paper is limited to a relatively qualitative
assessment of IMF-sensitive spectral features in the red part of the
spectra. The \nai\ doublet and the \wf\ Wing-Ford band are strong
in dwarfs and weak in giants, whereas the \cat\ triplet is weak
in dwarfs and strong in giants.
We find that all three features show considerable variation
within the sample. Na\,I and the Wing-Ford band vary by a factor
of $\sim 2$. 
When  abundance and age variations are ignored, this variation
directly translates into a variation of
a factor of $\sim 2$ in the number of low mass stars. 
Ca\,II varies 
only by $\sim 10$\,\%, but this
is expected as giants dominate the light.
As part of the analysis we demonstrate that the variation in
the Na\,I {\em feature} is indeed due to variation in the
strength of the Na\,I doublet and not driven by the neighboring
TiO bandhead (see, e.g., {Schiavon} {et~al.} 1997a, for a discussion of
this issue).\footnote{Although we can resolve this particular
issue, the Na\,I ambiguity illustrates the difficulty of
interpreting line indices: essentially
all indices reflect a canopy of blended spectral lines.}

The variation in IMF-sensitive features correlates with the velocity
dispersion of the galaxies: a higher velocity dispersion implies
stronger Na\,I, a stronger Wing-Ford band, and weaker Ca\,II.
The anti-correlation of Ca\,II and velocity dispersion was
previously discussed by {Cenarro} {et~al.} (2003), who also interpreted it as a
possible IMF effect. These results extend our earlier measurements
of very massive ellipticals in the Virgo cluster ({van Dokkum} \& {Conroy} 2010)
and metal-rich globular clusters in M31 ({van Dokkum} \& {Conroy} 2011); these previous
studies ``bookend'' the SAURON galaxies at very high and very low
dispersions respectively.

As shown in {Conroy} \& {van Dokkum} (2012) it is hazardous to derive quantitative
IMF constraints from these three features alone, as age and abundance
variations contribute to the observed absorption line strengths.
The Wing-Ford band is sensitive to the Fe abundance, Na\,I
is sensitive to [Na/Fe], and the Ca\,II triplet is
very sensitive to [Ca/Fe] (and the overall
$\alpha-$enhancement). All three indices also depend on
age, in complex ways (see, e.g., Fig.\ 12 in Conroy
\& van Dokkum 2012a). In our initial
paper on the most massive galaxies in Virgo and Coma
we mostly ignored these effects,
which was perhaps justified because the IMF
effects were so strong in that sample.
However, it is clear
that the trends in Fig.\ \ref{index.fig} to some extent reflect
the correlations of age and metal line abundances with
velocity dispersion
(see, e.g., {Trager} {et~al.} 2000a; {Thomas} {et~al.} 2005; {Kelson} {et~al.} 2006; {S{\'a}nchez-Bl{\'a}zquez} {et~al.} 2006; {Graves}, {Faber}, \& {Schiavon} 2009; {Scott} {et~al.} 2009; {Worthey}, {Ingermann}, \&  {Serven} 2011, and many other studies).

In a companion paper (Conroy \& van Dokkum 2012b) we use a comprehensive
stellar population synthesis model to quantify the IMF variation among
the early-type galaxies discussed in the present paper.
This model allows for abundance variations of
individual elements, which is critical as it removes the
ad-hoc assumption that we understand relative elemental
abundances better than we understand the IMF. Furthermore, we fit
the entire spectrum of each galaxy rather than line indices,
which means that blended spectral lines are treated correctly.

\begin{acknowledgements}

The data presented herein were obtained at the W.~M.~Keck Observatory,
which is operated as a scientific partnership among the California
Institute of Technology, the University of California and the National
Aeronautics and Space Administration. The Observatory was made
possible by the generous financial support of the W.~M.~Keck
Foundation.  The authors wish to recognize and acknowledge the very
significant cultural role and reverence that the summit of Mauna Kea
has always had within the indigenous Hawaiian community.  We are most
fortunate to have the opportunity to conduct observations from this
mountain. 

\end{acknowledgements}

\begin{appendix}

\section{Comparison to Van Dokkum (2008)}
Building on many previous studies of the fundamental plane
(e.g., {Djorgovski} \& {Davis} 1987; {van der Wel} {et~al.} 2004; {van Dokkum} \& {van der Marel} 2007) and the color-magnitude relation
(e.g., {Bower}, {Lucey}, \&  {Ellis} 1992a; {Stanford}, {Eisenhardt}, \&  {Dickinson} 1998; {Holden} {et~al.} 2004),
{van Dokkum} (2008) [vD08] constrained the slope
of the IMF near 1\,M$_{\odot}$ in early-type galaxies
by comparing their luminosity evolution
to their color evolution. This test, first
proposed by {Tinsley} (1980), is based on the expectation that
luminosity and color evolution depend on the IMF in different ways.
As discussed in {Tinsley} (1980) and in
\S\,2.1 of vD08 a more bottom-heavy
IMF should lead to slower luminosity evolution 
and faster color evolution. Therefore, a comparison
of luminosity evolution
to color evolution of a sample of galaxies should provide strong
constraints on the slope of the IMF near the main sequence turn-off
($\approx 1$\,\msun).

\begin{figure*}[htbp]
\epsfxsize=16.5cm
\epsffile{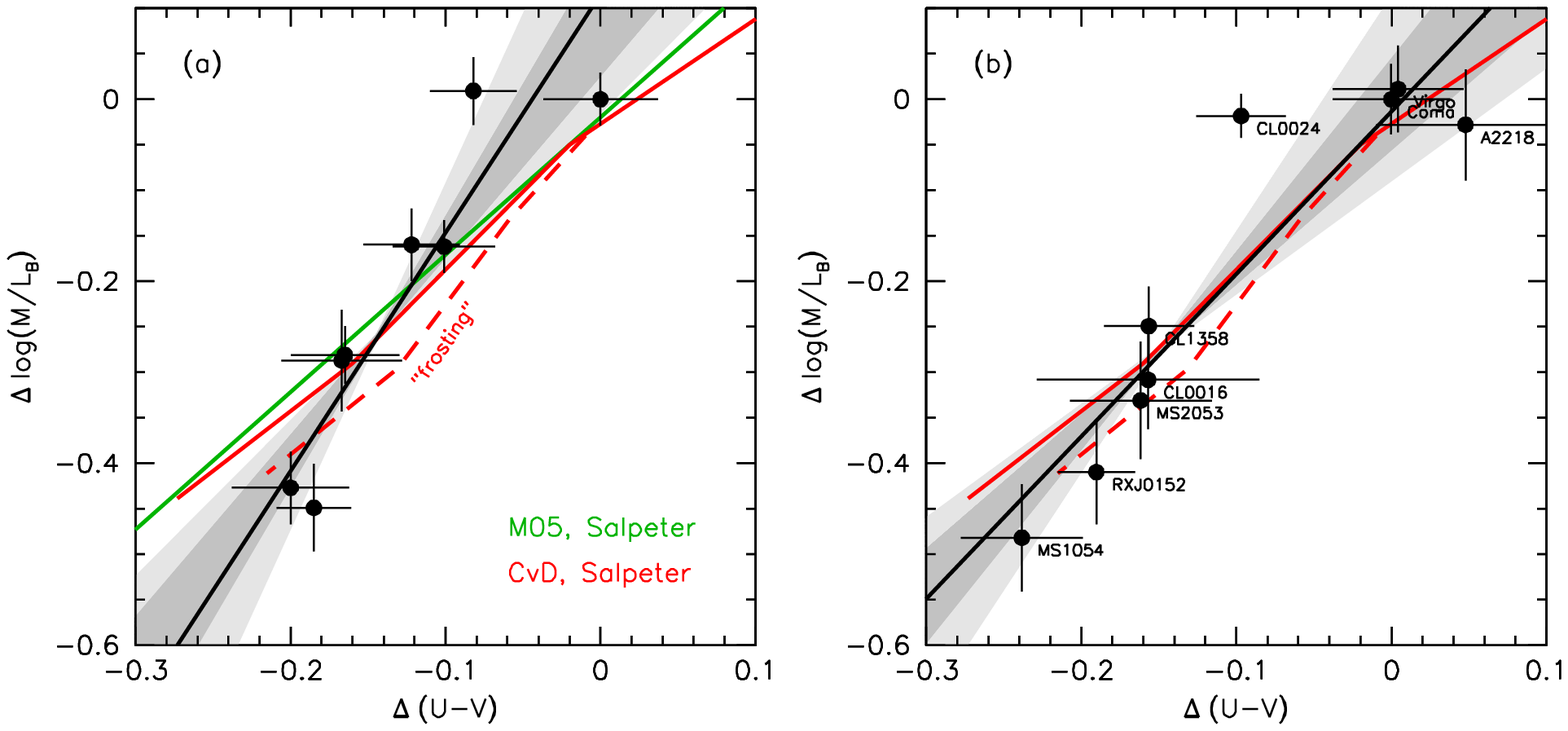}
\caption{\small
Color and luminosity evolution of early-type galaxies in clusters.
{\em (a)} Evolution at fixed dynamical mass, for galaxies
with $M>10^{11}$\,\msun. This panel is nearly identical to Fig.\ 5 in
van Dokkum (2008); the only difference is that
no corrections for progenitor bias were applied. Lines show expectations
for a Salpeter (1955) IMF, for a high metallicity Maraston (2005) model with
[Z/H]\,=\,0.35 (green)
and for an $\alpha-$enhanced Conroy \& van Dokkum (2012a)
model with [Fe/H]\,=\,0 and
[$\alpha$/Fe]\,=\,0.2 (red). The black line is the best fit
to the data; grey regions indicate the 68\,\% and 95\,\% confidence
limits of the best-fitting slope. The models predict lower
luminosity evolution at fixed color evolution than observed, although
a CvD model with ``frosting'' of young stars comes close to the data.
{\em (b)} Evolution at fixed velocity dispersion, for galaxies with
$\sigma>200$\,\kms, and including the Virgo cluster.
The CvD models are a satisfactory fit to the data.
\label{app.fig}}
\end{figure*}

The application of this test to massive
early-type galaxies in clusters
at $0<z<1$ yielded a surprising result: the slow
rest-frame $U-V$ color evolution of the galaxies and  fast
evolution of their rest-frame $M/L_B$ ratios
seemed to imply an IMF that is deficient in low mass stars
(``bottom-light'') compared to the IMF in the Milky Way.
The key
result from vD08 is shown in Fig.\ \ref{app.fig}a. The green line
is the predicted evolution of a {Maraston} (2005) model with super-Solar
metallicity ([Z/H]\,=\,0.35) and a
Salpeter IMF. This model is not a good fit to the data: the green line
has a slope $a = \Delta \log(M/L_B) / \Delta(U-V) = 1.5$, whereas a fit
to the data gives $a = 2.6$. From Eq.\ 5 and 6 in vD08 it follows that
the slope of the IMF near 1\,\msun\ is in the range $0.1\lesssim x
\lesssim 1.3$ depending on the metallicity,
where $x=2.3$ is the value for both a Milky Way IMF
and a Salpeter IMF.\footnote{Note that in vD08 the IMF was defined such that
the Salpeter form corresponds to a slope of 1.35.}

This finding
is in apparent conflict with the results in {van Dokkum} \& {Conroy} (2010, 2011) and
paper II.
Contradictory results are not exactly uncommon in this
particular field
(examples can be found in the
review by {Bastian}, {Covey}, \& {Meyer} 2010).
However, in this case the contradiction is rather extreme (bottom-light
versus bottom-heavy with respect to the Milky Way)
and applies to the exact same galaxies (massive
early-type galaxies in clusters).\footnote{For those readers
who failed to notice this: the 2008 and 2010 studies also have the
same first author. $^{\dagger}$\\
$~~~~~\dagger$\,This is not a coincidence:
the 2010 paper was partly motivated by the desire
to confirm $^{\ddagger}$ the conclusions of the 2008 paper using a more direct
method.\\
$~~~~~\ddagger$\,This did not work out quite as expected.}
Note that the results of vD08 are {\em not} in conflict with recent
mass measurements of early-type galaxies ({Treu} {et~al.} 2010; {Cappellari} {et~al.} 2012; {Spiniello} {et~al.} 2012), as a bottom-light IMF and a bottom-heavy IMF can result in
very similar $M/L$ ratios.\footnote{For a bottom-light IMF the ``extra''
mass is not in the form of low mass stars
but is comprised of the remnants of high mass stars (white dwarfs, neutron
stars, and black holes).}

Here we update the data and models of vD08 and examine whether they
can be brought into agreement with the absorption line
studies that indicate heavy mass functions.
Compared to the analysis in vD08 the following changes were made:
\begin{enumerate}
\item The [Z/H]\,=\,0.35 {Maraston} (2005) model was replaced by
an [Fe/H]\,=\,0, [$\alpha$/Fe]\,=\,0.2 {Conroy} \& {van Dokkum} (2012)
[CvD] model. As shown in Fig.\
\ref{app.fig} the $M/L$ ratio in the
CvD model evolves slightly faster in the age range 5--9
Gyr, although the difference is small. It remains true that
stellar population synthesis models are in reasonable agreement on
the evolution of rest-frame optical colors and luminosities.
\item A model with mild ``frosting'' of young stars was created (dashed
line in Fig.\ \ref{app.fig}. Based on Lick indices
{Trager}, {Faber}, \& {Dressler} (2008) finds that early-type
galaxies in the Coma cluster
have relatively young absolute luminosity-weighted ages, similar to early-type
galaxies in the general field. A possible explanation is that early-type
galaxies have a small fraction of relatively young stars in addition
to a dominant old population (see, e.g., {Trager} {et~al.} 2000a). The dashed
line is for a model in which 80\,\% of the stars have ages between
3 and 13.5 Gyr and 20\,\% of the stars are 3\,Gyr old.
\item In vD08 there was only one nearby cluster, Coma, for which
both accurate $U-V$ colors and $M/L$ ratio measurements were available.
We added the Virgo cluster to have another datapoint in the
$\Delta \log(M/L_B) \sim \Delta(U-V) \sim 0$ region of Fig.\ \ref{app.fig}.
Effective radii and surface brightnesses were obtained from
{Burstein} {et~al.} (1987). The effective radii are in excellent agreement
with the data in {Cappellari} {et~al.} (2006). Surface brightnesses were
corrected from the average surface brightness within $r_e$ to
the surface brightness at $r_e$ and corrected
for cosmological surface brightness dimming. A distance
of 16.5\,Mpc was assumed, based on surface brightness
fluctuations measured in the ACS Virgo Cluster Survey ({Mei} {et~al.} 2007).
Velocity dispersions were obtained from {Davies} {et~al.} (1987), multiplied
by 0.95 to undo the aperture correction, and then corrected
to a $3\farcs 4$ diameter circular aperture at the distance of Coma
(see {J\o{}rgensen}, {Franx}, \&  {Kj\ae{}rgaard} 1995). The
$U-V$ colors were obtained from {Bower}, {Lucey}, \&  {Ellis} (1992b).

\item A key assumption in vD08 was that structural evolution of the
galaxies could be ignored. As discussed in {Holden} {et~al.} (2010) this
assumption is important: the measured color and luminosity evolution can
be different from the true evolution if the masses and sizes of
the galaxies change with time. Following earlier results
for field galaxies
(e.g., {Daddi} {et~al.} 2005; {Trujillo} {et~al.} 2006; {van Dokkum} {et~al.} 2008)
there is now evidence for size evolution in clusters,
such that cluster galaxies of a given mass were smaller
at higher redshifts ({van der Wel} {et~al.} 2008; {Strazzullo} {et~al.} 2010; {Raichoor} {et~al.} 2012).
Whether this applies to all clusters is still unclear,
as is the question whether the size evolution
is driven by infall from the field
or changes to individual
cluster galaxies (e.g., {van der Wel} {et~al.} 2009). 
However, it does suggest that structural evolution needs to
be considered.
To address this issue {Holden} {et~al.} (2010) measured the color- and $M/L$
evolution of early-type galaxies at fixed velocity dispersion rather
than mass, reasoning that the velocity dispersion is probably a more
stable parameter (see, e.g., {Bezanson} {et~al.} 2011). From a comparison
of the Coma cluster to a single
cluster at $z=0.83$ they found that the color and
$M/L$ evolution of galaxies at fixed dispersion
is only $2.3\sigma$ removed from expectations of a Salpeter IMF.
We now follow {Holden} {et~al.} (2010) and measure offsets
in color and $M/L$ ratio from the $U-V$ -- $\sigma$ and
$M/L_B$ -- $\sigma$ relations,\footnote{Note that in calculating
$M/L$ ratios we are still
assuming that homology is conserved, which is almost certainly
incorrect; see, for instance, {van Dokkum} {et~al.} (2010)
and {Buitrago} {et~al.} (2011).} for galaxies with
$\sigma>200$\,\kms. 

\end{enumerate}
The results of these updates are shown in Fig.\ \ref{app.fig}b.
The data
are now in much better agreement with a Salpeter IMF.
The best-fitting relation has a slope of $a = 1.81
\pm 0.27$, which means that the high-metallicity
{Maraston} (2005) model with a Salpeter 
IMF is only $1.2\sigma$ removed from the data. The {Conroy} \& {van Dokkum} (2012) model
is in even better agreement, particularly if some frosting is included.
We infer that the luminosity and color evolution of massive
early-type
galaxies does not rule out IMFs with Salper-like slopes near
$\sim 1$\,\msun, 
contrary to the conclusions of vD08.
\newpage

\end{appendix}

\bibliography{}


\end{document}